\documentclass[journal]{IEEEtran}
\usepackage{xcolor}
\usepackage{multirow}
\usepackage{filecontents}

\usepackage[pdftex]{graphicx}
\ifCLASSINFOpdf
\else
\fi
\usepackage{amssymb}
\usepackage{amsmath}
\usepackage{mathtools}
\usepackage{amsthm}
\usepackage[utf8]{inputenc}
\usepackage[english]{babel}

\usepackage{algorithmic}
\usepackage{soul}
\usepackage{dblfloatfix}
\usepackage{url}

\begin{document}
%
\title{A Configuration Based Pumped Storage Hydro Model in MISO Day-Ahead Market}
%
%
%
\author{Bing~Huang,~\IEEEmembership{Member,~IEEE,}
        Yonghong~Chen,~\IEEEmembership{Senior~Member,~IEEE}
        and~Ross~Baldick,~\IEEEmembership{Fellow,~IEEE} 

\thanks{B. Huang and Y. Chen are with Midcontinent Independent System Operator, Inc. (MISO), Carmel, IN, 46032 USA.}
\thanks{R. Baldick is an emeritus professor at the Department
of Electrical and Computer Engineering, University of Texas, Austin, TX, 78712 USA.}
\thanks{This work has been submitted to the IEEE for possible publication. Copyright may be transferred without notice, after which this version may no longer be accessible.}
}

\maketitle
\begin{abstract}

Pumped storage hydro units (PSHU) can provide flexibility to power systems. This becomes particularly valuable in recent years with the increasing shares of intermittent renewable resources. However, due to emphasis on thermal generation in the current market practices, the flexibility from PSHUs have not been fully explored and utilized. This paper proposes a configuration based pumped storage hydro (PSH) model for the day-ahead market, in order to enhance the use of PSH resources in the system. A strategic design of incorporating and fully optimizing PSHUs in the day-ahead market is presented. We show the compactness of the proposed model. Numerical studies are presented in an illustrative test system and the Midcontinent Independent System Operator (MISO) system.


\end{abstract}

\begin{IEEEkeywords}
Pumped storage hydro, configuration based model, mixed-integer programming, security constrained unit commitment.
\end{IEEEkeywords}
%
\IEEEpeerreviewmaketitle

\section*{Nomenclature}

Sets and indices:


\begin{IEEEdescription}[\IEEEusemathlabelsep\IEEEsetlabelwidth{$V_1,V_2.5$}]
	\setlength{\itemsep}{2pt}
	\item[$t \in \mathcal{T}$] set of time intervals;
	\item[$g\in \mathcal{G}_{psh}$] set of PSHUs;
	\item[$g\in \mathcal{G}_{psh,r}$] set of PSHUs that share the same reservoir $r$;
	\item[$g\in \mathcal{G}$] set of the rest of the generating units in a system;
	\item[$m\in \mathcal{M}_{g}$] set of configurations, $\mathcal{M}_{g}=[\textit{alloff}, gen, pump]$;
	\item[$n\in \mathcal{M}_{g}^{F,m}$] set of configurations that configuration $m$ can feasibly transit to;
	\item[$r \in \mathcal{R}$] set of reservoirs.
\end{IEEEdescription}

Data [units]:




\begin{IEEEdescription}[\IEEEusemathlabelsep\IEEEsetlabelwidth{$V_1,V_2.5$}]
	\setlength{\itemsep}{2pt}
	\item[$D_t$] system net load at period $t$ [\$/MW];
	\item[$\underline{Q}_{g}^{gen}$] minimum generation power of PSHU $g$ [MW];
	\item[$\overline{Q}_{g}^{gen}$] maximum generation power of PSHU $g$ [MW];
	\item[$\underline{Q}_{g}^{pump}$] minimum pumping power of PSHU $g$ [MW]; 
	\item[$\overline{Q}_{g}^{pump}$] maximum pumping power of PSHU $g$ [MW];
	\item[$\eta_g^{gen}$] generating efficiency of the PSHU $g$ [NA];
	\item[$\eta_g^{pump}$] pumping efficiency of the PSHU $g$ [NA];
	\item[$E_{r,0}$] initial energy level of the reservoir $r$ [MWh];
	\item[$E_{r,|\mathcal{T}|}$] final energy level of the reservoir $r$ [MWh];
	\item[$\overline{E}_r$] maximum energy level of the reservoir $r$ [MWh];
	\item[$\underline{E}_r$] minimum energy level of the reservoir $r$ [MWh];
	\item[$C_{g,t}^{pump}$] the bid price of pump load at unit $g$ during time interval $t$ [\$/MW]. 
\end{IEEEdescription}

Variables [units]:

\begin{IEEEdescription}[\IEEEusemathlabelsep\IEEEsetlabelwidth{$V_1,V_2.5$}]
	\setlength{\itemsep}{2pt}
	\item[$e_{r,t}$] energy stored in the reservoir $r$ at time $t$ [MWh];
	\item[$u_{g,t}^{m}$] binary variable, commitment variable of unit $g$ configuration $m$ during time interval $t$ [NA];
	\item[$ur^{m}_{r,t}$] continuous variable, if $ur^{m}_{r,t}=1$, it represents the status of reservoir $r$ in mode $m \in \{gen,pump\}$, at time interval $t$ [NA]; 
	\item[$v_{g,t}^{m,n}$] binary variable, transition variable between configuration $m$ and configuration $n$ of PSHU $g$ during time interval $t$  [NA];
	\item[$q_{g,t}^{gen}$] continuous variable, amount of generation at a PSHU $g$ during time interval $t$ [MW];
	\item[$q_{g,t}^{pump}$] continuous variable, amount of pumping load at a PSHU $g$ during time interval $t$ [MW];
	\item[$q_{g,t}$] continuous variable, amount of generation at unit $g$ during time interval $t$ [MW].
	
\end{IEEEdescription} 


	\setlength{\itemsep}{2pt}
	

Auxiliary Variables [units]:

\begin{IEEEdescription}[\IEEEusemathlabelsep\IEEEsetlabelwidth{$V_1,V_2.5$}]
	\setlength{\itemsep}{2pt}
	\item[$f_{g,t}^{gen}$] continuous variable, energy opportunity cost of $gen$ configuration offered at PSHU $g$ during time interval $t$ [\$ /hr];
	\item[$C(q_{g,t})$] cost function of generating unit $g$ [\$ /hr].
\end{IEEEdescription}

\section{Introduction}
\subsection{Background and Motivation}
%
%
%
%

\IEEEPARstart{P}{umped} storage hydro units (PSHU) could provide additional flexibility to power systems. This flexibility is critical to enhance the reliability of modern power systems that are experiencing more uncertainties nowadays. There are a total of $40$ pumped storage hydro (PSH) plants with over $22$ GW capacity in the United States \cite{eiapumped2013}, which roughly equals $2
$\% of U.S. generating capacity \cite{eiapumped2013}. In the 2016 Hydropower Vision Report (DOE Report)  \cite{DOE2016}, the investigation explores a range of growth scenarios, finding that the
pumped storage capacity can increase in both the near
term (2030), by 16.2 GW, and in the longer term (2050), by
an additional 19.3 GW, for a total of 35.5 GW deployed by
2050 \cite{DOE2016} (pp 17-19). However, the footprint and required topography for reservoirs pose significant regulatory challenges such as licensing and satisfying environmental requirements. 


One of the major benefits a PSHU brings to the system is it stockpiles excess electricity that enables base-load generators to stay on line when the net load is low. The number of expensive shut down and start up operations can be reduced at those generators. The energy stored in a PSH reservoir is used to generate electricity
later when it is needed. Therefore, the system overall operation cost is reduced. 

This load/generation-shifting effects are particularly useful in the current system with greater uncertainties on both generation and (net) load sides. In addition, a PSHU is able to provide a variety of services ranging from weekly and daily smoothing of load to providing hourly and subhourly reserves to responding to frequency control within a minute \cite{barton2004energy}. Such flexibility provided by PSHUs would significantly facilitate the integration of renewable resource with lower costs. 

However, these important and valuable services that are available from PSHUs have not been utilized largely due to the fact that PSHUs have not been fully optimized in the market. For example, the owners of PSHUs on the footprint of Midcontinent Independent System Operator (MISO) make their decisions of pumping and generating based on their forecast of market prices.

This practice of PSH technology has two drawbacks: First, as a market participant with limited information about the market, the forecast of market prices can deviate significantly from the realization. Therefore, the decisions made based on the forecast would impair profits for the PSHU in short term and this would discourage the development and investment in the PSHU in long term; Secondly, the decisions made by PSHU are sub-optimal to the system welfare. That is, the benefits and the flexibility provided by PSHU are not fully exploited by the system under the current practice.

To overcome the drawbacks of the current practice, introducing the PSHUs into MISO's day-ahead unit commitment model is a first step. Therefore, a suitable model for PSHU in a unit commitment (UC) problem is studied in this paper.

\subsection{Literature Review}

One of the important benefits a PSHU can provide to a system is smoothing power injections from renewable resources; therefore, PSHU models have been widely discussed and deeply developed for operation with renewable resources. PSH models have also been studied in unit commitment problems and stand alone systems.  

In \cite{castronuovo2004optimization} and \cite{duque2011optimal}, a profit maximization problem is formulated for a single wind farm together with a PSH plant. A hybrid wind plus general storage plant model is presented in \cite{korpaas2003operation}. The impact of energy storage sizing on wind-hydro system operation and economics has been studied in \cite{korpaas2003operation}, \cite{castronuovo2004optimal} and \cite{brown2008optimization}. In these works, a PSH plant is modeled as two individual units: a pump and a generator. The state of charge of the storage is modeled with energy balance constraints across each pair of successive time intervals with efficiency lost.

In \cite{brown2008optimization}, additional constraints are introduced to meet spinning reserve requirements or frequency regulation unit commitment requirements. 
The bidding and scheduling of a PSHU is studied as part of a generating company with hydro, thermal and PSHUs in \cite{ni2004optimal}. 
In \cite{garcia2008stochastic}, stochastic joint optimization is used to maximize the profit of wind generation and PSHUs in an electricity market. 

 


In \cite{castronuovo2004optimization}-\cite{garcia2008stochastic}, the market prices are taken as an input for the profit maximization problem at a renewable-pumped storage hydro hybrid plant level. The feasibility of standalone solar-pumped storage hydro and solar-wind-pumped storage hydro is studied in an island mode by \cite{ma2015pumped} and \cite{ma2014technical}. 
 
Although these works provide a variety of models for a PSHU, incorporating a PSHU in an electricity market clearing process is not their major concern. The following paragraphs discuss formulations that are more suitable for incorporation in a market clearing process.

A robust unit commitment problem with wind power and a PSHU is studied in \cite{jiang2012robust}. A binary variable is introduced to indicate whether the unit absorbs (pumping) or generates electricity. In this model, a PSHU has to be either pumping or generating at a given time. However, for some PSH plants, due to their physical or operational limits, there are minimum outputs for each mode. That is, because of the minimum limits, the PSHUs are either generating or pumping at least at their minimum limits. However, at certain times in the day it may be the best to neither pump or generate. In other word, constantly charging or discharging the PSHU is not always the best strategy in the system operation. 


The introduction of an idle mode would allow a PSHU to turn off. The scheduling of PSHUs are modeled in a day-ahead market in \cite{khodayar2013enhancing} and \cite{li2016enhanced}. Three modes, pumping, generating and idle are modeled for each of the pump storage units. However, the transitions between each pair of modes are not specifically modeled in these works. Therefore, some operation features, for example those related to transition time between modes of a PSHU, are either simplified or ignored.

Some of the operation features of a combined cycle gas turbine unit in a day-ahead market are similar to those of a PSHU. In particular, we found the configuration based model used for combined cycle gas turbines (CCGTs) in unit commitment problem described in \cite{chen2017mip}, \cite{blevins2007combined} and \cite{hui2011combined} fits well with the operation of a PSHU. Similar to CCGTs, a PSHU can also be modeled with different operational configurations such as pumping and generating. An ``alloff'' configuration can be introduced to allow the unit to stay idle. 





\subsection{Contributions}

This paper focuses on developing a new model for a PSHU in day-ahead market. The contributions are summarized below:

\begin{itemize}
    \item A configuration based PSHU model is developed in a unit commitment problem such that the flexibility of existing pumped storage hydro plants can be leveraged at the system level. 
    \item An ``Alloff'' configuration is introduced to allow the unit to stay idle at anytime within the operating window. Conveniently, offline supplemental reserve can be cleared for the ``Alloff'' configuration of a PSHU. What is more, transitions between each pair of modes are modeled specifically. The state of charge constraints are modeled for each reservoir included sharing of reservoirs by multiple units.
    \item The compactness of the model is compared with configuration based combined cycle gas turbine models. 

    \item The proposed model is demonstrated in a test system and implemented in a MISO system with a benefit analysis and a computational study.  
    
\end{itemize}

Notice that it is necessary to specifically model the transitions between different modes. The inter-temporal constraints related to the transitions such as min up/down time and transition time in a PSHU varies and can be accurately modeled. For example, the transition time needed from a generation mode to a pump mode may different from the transition time from an alloff mode to a pump mode. What is more, it is recognized that there is a fixed generation output pattern during a transition from one configuration to another in a combined cycle unit \cite{hua2019tight}. A similar pattern exists in the operation of some PSHUs. Having the transitions between different modes explicitly modeled enables the representation of these operational features and would make the model flexible and easy to extend into a market with various length of time intervals.  

The rest of the paper is organized as follows. Section \ref{sec:model} presents the proposed Mixed Integer Programming (MIP) formulation for PSHUs. The configuration based formulation is introduced in details and the compactness of the model is discussed. Section \ref{sec:case_studies} gives numerical study results on a small test system and the large MISO system. A detailed benefit analysis is presented with MISO case studies. Section \ref{sec:conclusion} concludes this paper.

\section{Configuration-Based Modeling of a Pumped Storage Hydro Unit}
\label{sec:model}
\subsection{Mathematical Formulation}
\begin{figure}[!h]
	\centering
	\includegraphics[width=0.42\textwidth]{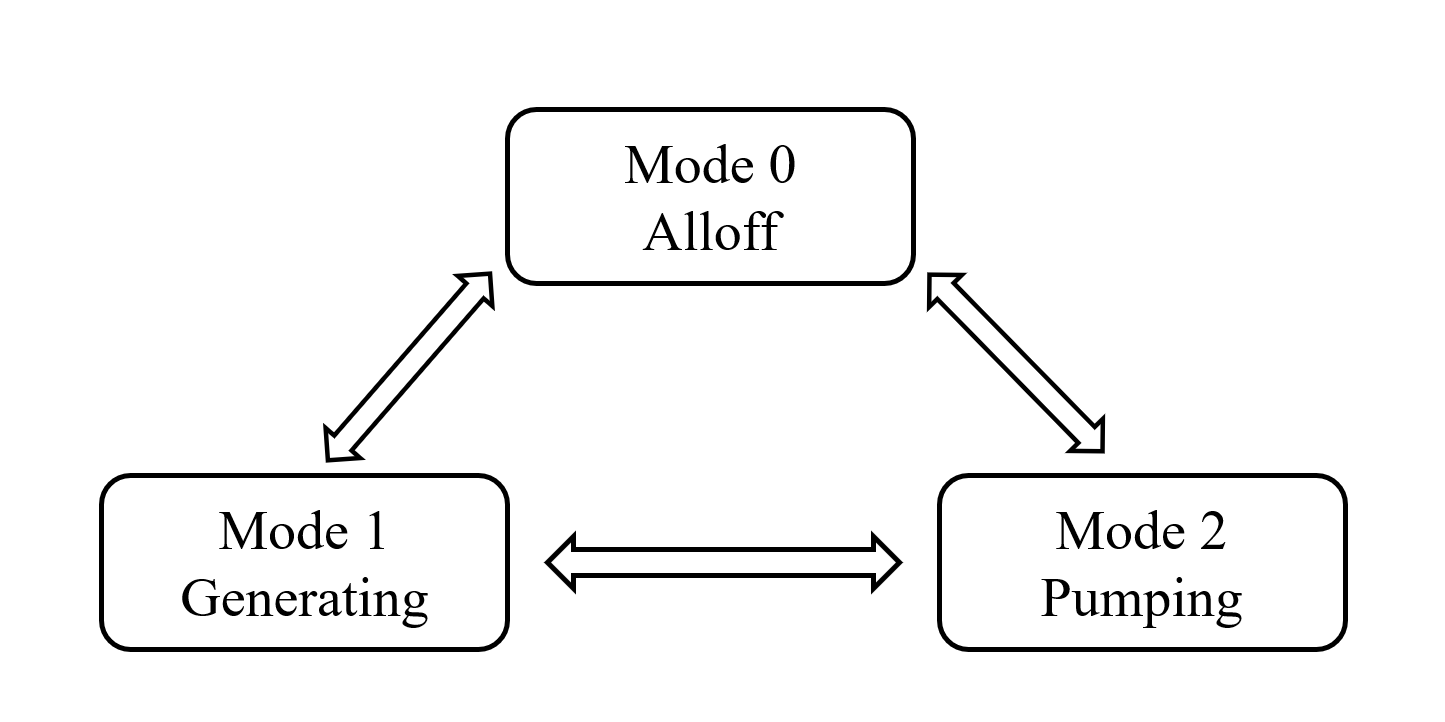}
	\caption{Mode transition diagram of a PSHU in two consecutive periods.}
	\label{fig:state_gram}
\end{figure}

The configuration based modeling of PSHU represents all feasible operation modes of a PSHU. A pumped storage hydro plant can contain multiple units and each of them will be modeled individually. There are only three operation modes in a PSHU namely generating, pumping and offline. Transitions are allowed between each pair of these modes shown in Fig. \ref{fig:state_gram}. The ``Mode $0$'' represents the state when the unit is offline. 

\subsubsection{Objective Function}
The objective of the unit commitment problem is to minimize the system operating costs. Under MISO current practice, the costs related to a PSHU is the offered production costs of the generating mode minus the bid prices of the pumping mode which is reflected as negative costs in (\ref{eq:objective}). These terms in box \eqref{eq:objective} will be removed from the objective ultimately because the operating cost of PSH is close to zero, but are presented here for the discussion in Section \ref{sec:case_studies}. The third term in (\ref{eq:objective}) represents the piece-wise linear production costs of the rest of generators in the system. 
\begin{equation} 
\underset{q,u}{\min}\boxed{\sum_{g \in \mathcal{G}_{psh}} \sum_{t \in \mathcal{T}}(f_{g,t}^{gen} - C_{g,t}^{pump} q_{g,t}^{pump})} + \sum_{g \in \mathcal{G}} \sum_{t \in \mathcal{T}}C(q_{g,t}). 
\label{eq:objective}
\end{equation}

\subsubsection{System Energy Balance Constraints}
The generation has to be balanced with net load in the system at all times. In (\ref{eq:energy balance}), during each interval $t$, the total generation in the system including the generation from PSHUs on the left should be balanced with the sum of the net load and the pumping load from the PSHUs on the right. 
\begin{equation}
\sum_{g\in \mathcal{G}}q_{g,t} + \sum_{g\in \mathcal{G}_{psh}}q_{g,t}^{gen} = D_t + \sum_{g\in \mathcal{G}_{psh}}q_{g,t}^{pump},\ \forall t\in \mathcal{T}. 
\label{eq:energy balance}
\end{equation}

\subsubsection{State and Transition Logic Constraints}

Constraints (\ref{eq:mutual}) guarantee that the unit commitment variables of each mode in a PSHU described in Fig. \ref{fig:state_gram} are mutually exclusive, as modeled for CCGTs in \cite{liu2009component}:
   
\begin{equation}
\begin{split}
\sum_{m\in \mathcal{M}_g}{u_{g,t}^{m}} = 1,\ \ \forall g \in \mathcal{G}_{psh}, \forall t \in \mathcal{T}. 
\end{split}
\label{eq:mutual}
\end{equation}

The transition between two modes $m,n$ in a PSHU $g$ at time $t$ is defined as a  binary variable $v_{g,t}^{m,n}$. Notice that the start up and shut down of a mode are modeled as the transition between the mode and the \textit{alloff} mode. These constraints are modeled for CCGTs in \cite{morales2016tight}.   

\begin{equation}
\begin{split}
\ u_{g,t}^m - u_{g,t-1}^m =& \sum_{n\in \mathcal{M}_g^{F,m}}{v_{g,t}^{n,m}} - \sum_{n\in \mathcal{M}_g^{F,m}}{v_{g,t}^{m,n}},\ \\& \forall g \in \mathcal{G}_{psh}, \forall m \in \mathcal{M}, \forall t \in \mathcal{T}. 
\end{split}
\end{equation}

In addition to the mutual exclusivity constraints on the commitment variable of each configuration, there should be at most one feasible transition at any time \cite{chen2017mip}:

\begin{equation}
\label{eq:transit_mutual}
\begin{split}
\sum_{m\in \mathcal{M}_{g}}\sum_{n\in \mathcal{M}_g^{F,m}}{v_{g,t}^{m,n}}\leq 1, \ \ \forall g \in \mathcal{G}_{psh}, \forall t \in \mathcal{T}. 
\end{split}
\end{equation}

\subsubsection{Box constraints}
The amount of pumping load during interval $t$ from the PSHU is constrained by the capacity of the reservoir and the capacity of the pump unit in \eqref{eq:pumpbox}. The pump output of a PSHU will be forced to zero by \eqref{eq:pumpbox} when $u_{g,t}^{pump}=0$ indicating the unit is not in a pumping mode. Symmetrically, the amount of generation during interval $t$ from the PSHU is constrained by the capacity of the reservoir and the capacity of the generation unit shown in (\ref{eq:hydrobox}). The generation output of a PSHU will be forced to zero by \eqref{eq:hydrobox} when $u_{g,t}^{gen}=0$ indicating the unit is not in a generating mode. 
\begin{equation}
\begin{split}
u_{g,t}^{pump}\underline{Q}_g^{pump} \leq q_{g,t}^{pump} \leq u_{g,t}^{pump} \min\{\frac{\overline{E}_r-\underline{E}_r}{\eta_g^{pump}},\overline{Q}_g^{pump}\}, \\ \forall r \in \mathcal{R}, \forall g \in \mathcal{G}_{psh,r}, \ \forall t \in \mathcal{T}.
\label{eq:pumpbox}
\end{split}
\end{equation}
\begin{equation}
\begin{split}
	u_{g,t}^{gen}\underline{Q}_g^{gen}  \leq q_{g,t}^{gen} \leq u_{g,t}^{gen} \min\{(\overline{E}_r-\underline{E}_r)\eta_g^{gen},\overline{Q}_g^{gen}\},\\ \forall r \in \mathcal{R}, \forall g \in \mathcal{G}_{psh,r}, \ \forall t \in \mathcal{T}.
	\label{eq:hydrobox}
\end{split}
\end{equation}

\subsubsection{Storage Energy Balance and State of Charge (SOC) Constraints}

The energy stored in the PSH system is linked at each consecutive time interval as shown in (\ref{eq:StorageEnergyBal}). Notice that there can be more than one PSHU sharing a reservoir in the model. Parameters $\eta_g^{gen}$ and $\eta_g^{pump}$ are the efficiencies of generating and pumping indicating energy loss in both modes. The energy stored in the reservoir at the beginning and end of the day is given by (\ref{eq:socb}) and (\ref{eq:soce}), respectively. The upper and lower bounds of the SOC are provided by (\ref{eq:soct}). The state of charge constraints modeled in this paper are based on previous work \cite{castronuovo2004optimization}-\cite{garcia2008stochastic}. 
\begin{equation}
\begin{split}
e_{r,t+1}=e_{r,t} + \sum_{g\in \mathcal{G}_{psh,r}} \eta_g^{pump}q_{g,t}^{pump}-\sum_{g\in \mathcal{G}_{psh,r}} \frac{q_{g,t}^{gen}}{\eta_g^{gen}}, \\  \forall r \in \mathcal{R}, \ \ \forall t \in [0,|\mathcal{T}|-1].
\end{split}
\label{eq:StorageEnergyBal} 
\end{equation}

\begin{equation} e_{r,0}=E_{r,0}, \ \ \forall r \in \mathcal{R}. 
\label{eq:socb}
\end{equation}

\begin{equation} e_{r,|\cal{T}|}=E_{r,|\mathcal{T}|}, \ \ \forall r \in \mathcal{R}.
\label{eq:soce}
\end{equation}

\begin{equation} \underline{E_r} \leq e_{r,t} \leq \overline{E_r}, \ \ \forall r \in \mathcal{R}, \ \ \forall t \in \mathcal{T}.
\label{eq:soct}
\end{equation}

The start up and shut down time, transition time, minimum up/down time and security constraints are not listed here. They can be easily handled in the configuration based model. The details can be found in \cite{chen2017mip}.

\subsubsection{Practical Operational Limits} To demonstrate the adaptability of the proposed configuration based PSH model to industry practice, two additional constraints are presented to reflect some of the physical limits the PSHUs have in their daily operations.

\begin{equation}
\label{eq:pump_startup_limit}
\begin{split}
\sum_{g\in \mathcal{G}_{psh,r}}\sum_{n\in \mathcal{M}_g^{F,pump}}{v_{g,t}^{n,pump}}\leq N, \\ \forall r \in \mathcal{R}, \forall t \in \mathcal{T}. 
\end{split}
\end{equation}

At some PSH plant, due to the physical limits in the start up procedure of pump units, only a limited number of pump units can be brought online in a given time period. In constraints \eqref{eq:pump_startup_limit}, $\mathcal{M}_g^{F,pump}$ is the set of modes for which unit $g$ can feasibly transit to a \textit{pump} mode, bearing in mind that $v_{g,t}^{n,pump}$ is the transition variable of unit $g$ from mode $n$ to the \textit{pump} mode. Therefore, without introducing new variables, constraints \eqref{eq:pump_startup_limit} precisely capture the operational feature that no more than $N$ units sharing reservoir $r$ can transit from any mode to a pumping mode in time interval $t$. 

For the PSH plant with large reservoirs, there are typically multiple PSH units installed in the plant and they are jointly operated with the reservoirs. It is not economical and physically not feasible for the plant to have one unit pumping and another generating at the same time. To incorporate this feature for a PSH plant with multiple units, constraints \eqref{eq:mutually_exclusive_plant_level0} and \eqref{eq:mutually_exclusive_plant_level1} are introduced.

\begin{equation}
\label{eq:mutually_exclusive_plant_level0}
\begin{split}
ur^{pump}_{r,t}+ur^{gen}_{r,t} \leq 1, \ \ \forall r \in \mathcal{R}, \forall t \in \mathcal{T}.
\end{split}
\end{equation}

\begin{equation}
\label{eq:mutually_exclusive_plant_level1}
\begin{split}
u^{m}_{g,t}\leq ur^{m}_{r,t}, \ \  \forall r \in \mathcal{R}, \forall g \in \mathcal{G}_{psh,r}, \\ \forall m \in \{gen,pump\}, \forall t \in \mathcal{T}.
\end{split}
\end{equation}

A pair of variables $ur^{pump}_{r,t}, ur^{gen}_{r,t}$ are introduced for a reservoir or a PSH plant $r$ to represent the status of the plant as pumping or generating at time interval $t$. Therefore, constraints \eqref{eq:mutually_exclusive_plant_level0} are the mutual exclusivity constraints at the plant level with \eqref{eq:mutually_exclusive_plant_level1} constraining $u^m_{g,t}$ which is the commitment variable of PSHU $g$ in mode $m$ at time interval $t$. Constraint \eqref{eq:mutually_exclusive_plant_level1} indicates if any unit $g$ of the plant $r$ is in \textit{pump} mode then the plant status will be in \textit{pump} mode indicated by $ur^{pump}_{r,t}=1$. The same for the \textit{gen} mode. Notice that since $u^m_{g,t}$ is binary, $ur^m_{r,t}$ can be continuous and bounded by \eqref{eq:mutually_exclusive_plant_level0} and \eqref{eq:mutually_exclusive_plant_level1}.

Combining \eqref{eq:mutually_exclusive_plant_level0} and \eqref{eq:mutually_exclusive_plant_level1}, if any unit in a reservoir is generating at a time interval, all the other units sharing the same reservoir would not pump at the same time interval and vice versa.

\subsection{Compactness of the Proposed Formulation}
This section analyzes the scale of the proposed model in comparison with the configuration based combined cycle gas turbine model in \cite{chen2017mip} and \cite{morales2016tight}. The proposed model is formulated via constraints \eqref{eq:mutual}-\eqref{eq:mutually_exclusive_plant_level1}. Compared to the combined cycle model in \cite{chen2017mip} and \cite{morales2016tight}, storage related constraints \eqref{eq:StorageEnergyBal}-\eqref{eq:mutually_exclusive_plant_level1} represent the major differences. The impacts of the storage related constraints to computational complexity of the problem is analyzed with following parameters:


\begin{itemize}
    \item Number of Variables: There is only one continuous variable at each time interval for a reservoir. Two additional continuous variables are needed for each reservoir to enforce the mutual exclusivity constraints at the reservoir level if there are multiple PSHUs operating in the same reservoir.
    \item Number of Constraints: The increases in the number of constraints on storage energy balance \eqref{eq:StorageEnergyBal}, state of charge \eqref{eq:soct}, the pump start up constraints \eqref{eq:pump_startup_limit} and plant level mutual exclusivity constraints \eqref{eq:mutually_exclusive_plant_level0}-\eqref{eq:mutually_exclusive_plant_level1} are proportional to the number of reservoirs and the number of time intervals.
    \item Number of Non-zeros: There are only a few non-zeros added by each constraint of \eqref{eq:socb}-\eqref{eq:mutually_exclusive_plant_level1}. However, the inter-temporal storage energy balance constraint \eqref{eq:StorageEnergyBal} would introduce more non-zeros depending on the number of PSHUs sharing the reservoir.  
\end{itemize}

The impacts of the storage related constraints on problem computational complexity are moderate except for the storage energy balance constraints \eqref{eq:StorageEnergyBal}. Particularly, because it is common that multiple units share a reservoir in a PSHU, the increase of non-zeros per constraint can be significant.     

\section{Numerical Results}
\label{sec:case_studies}

In this section, first, the current PSHU model at MISO is introduced and compared with the proposed model in a case study with two PSHUs. Then, the proposed model is applied to a MISO system. Benefit analyses and computational studies are presented.  

\subsection{Two Units Case Study}
\label{sec:case_studies_2units}
A day ahead unit commitment (UC) and economic dispatch (ED) problem formulated in \eqref{eq:objective} - \eqref{eq:soct} is solved with a $24$ hours net-load scenario. The results from current MISO model and proposed model will be compared. For the sake of a simple and clear illustration, the practical operational limits in \eqref{eq:pump_startup_limit}-\eqref{eq:mutually_exclusive_plant_level1} are not included, minimum up/down time, reserve requirements, ramp constraints, and transmission security constraints are ignored in this small case study.

Table \ref{table:psh_unit} shows the units considered in this case study. The two pumped-storage hydro units that share a reservoir have the same parameters for both Pump and Gen modes as listed once in the table. Three thermal generators with different capacities and different costs are included to represent the generations besides pumped-storage hydro units in the system. For a simple presentation, the marginal generation costs and bid prices are constants over feasible generation and consumption levels and independent of time for all units. Notice that pumping in both units are ``block loaded'' meaning that the pumping load is either at a predetermined level or zero. This is a typical operating feature of the pumped-storage hydro units in the MISO system. The energy efficiency of the pumping and generating processes are identical in both units. 


\begin{table}[!h]
	\caption{Unit Parameters}
	\label{table:psh_unit}
	\centering
	\begin{tabular}{l c c c c c}
		\hline
		\multirow{2}{*}{Unit}  & Cost/Price & $\underline{q}^m$ & $\overline{q}^m$ & $\eta_g^m$  \\
		& \$/MWh & MW & MW &  \\
		\hline
		PSHU Pump & $24$ & $200$ & $200$ & $0.9$\\
		PSHU Gen  & $26$ & $100$ & $200$ & $0.9$\\
		Thermal Gen 1 & $30$& $0$ & $600$ & NA\\
		Thermal Gen 2 & $20$ & $0$ & $400$ & NA\\
		Thermal Gen 3 & $15$ & $0$ & $500$ & NA\\
		\hline
	\end{tabular}
\end{table}

The energy price at each node in the transmission network or locational marginal price (LMP) is the dual value at the energy balance constraint in \eqref{eq:energy balance} after the problem been solved. Notice that since there is no transmission network constraint in this case study, there is a single marginal price for the whole system at every time interval. 

\begin{figure}
	\centering
	\includegraphics[width=\linewidth]{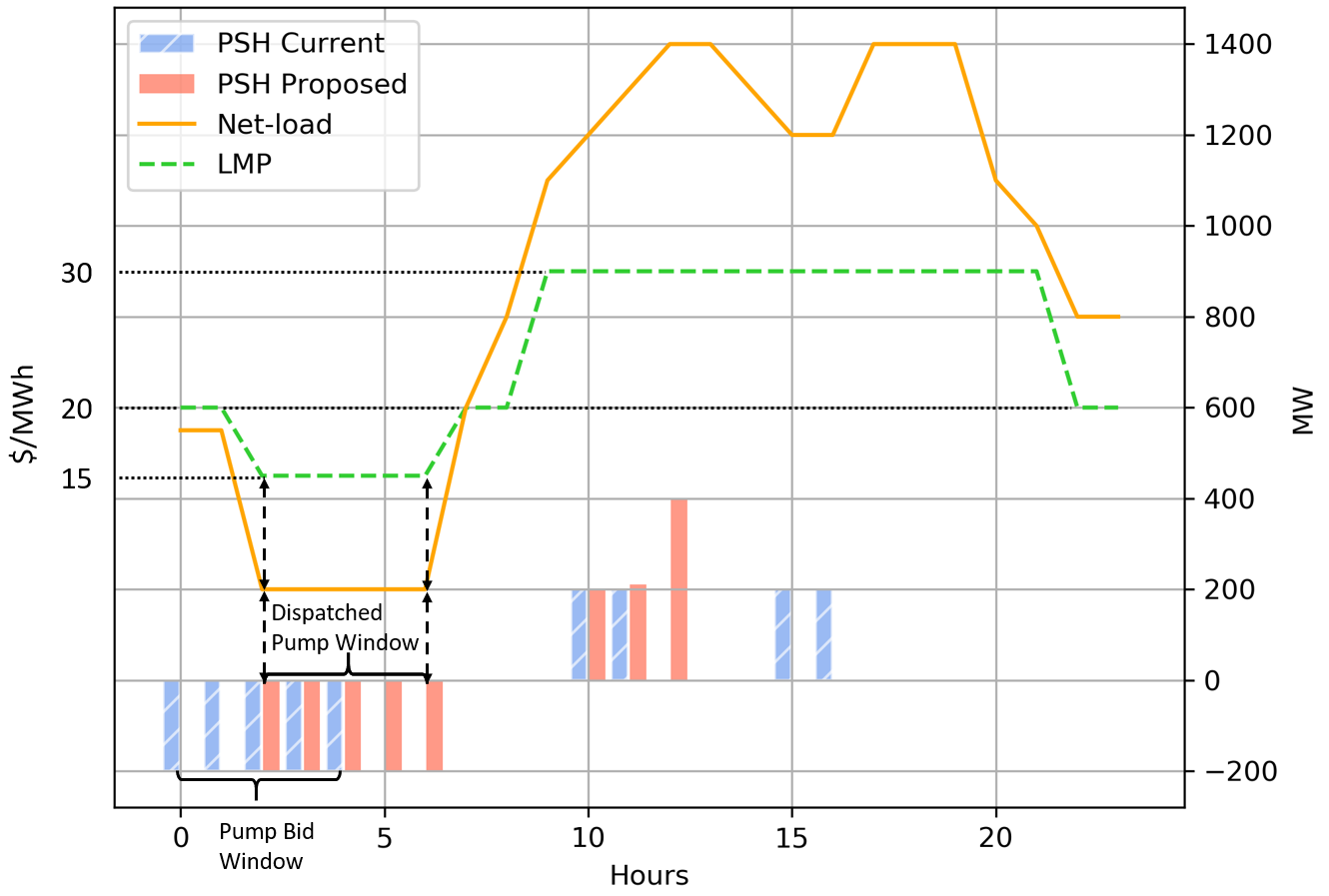}
	\caption{Small Case Study: PSHU Dispatch Results.}
	\label{fig:two_units_case_study}
\end{figure}

\subsubsection{Current PSHU Model at MISO}\label{cha:PSHU-current-practice}

In the current MISO day ahead market, PSHUs offer opportunity costs and bid prices for their generation and pump mode, respectively. State of charge limits for their reservoirs are not enforced explicitly by the system operator. Instead, a maximum daily electricity generation limit is submitted and applied to PSHUs for their generation modes. The PSHU owner determines the pump/generate window. For example, the owner may bid one of their units as pumping load only between $0-4$ AM and that generation can occur during the rest of the day. 

The PSHU result using the current MISO model is compared with the results from the proposed model in Fig. \ref{fig:two_units_case_study}. The PSHU outputs from the current model are represented by shaded blue bars and the PSHU outputs from the proposed model are represented by red solid bars. Notice that net-load $D_t$ and PSH output are aligned with the axis on the right in units of MW while LMP aligns with the axis on the left in units of \$/MWh. The negative bars indicate the pumping load of the PSHU and the positive bars indicate the generation of the PSHU. Net-load in this paper is defined as the system fixed demand minus the renewable generations not including pumped-storage hydro units. 

Although state of charge are not represented in the market clearing process in the current model, it is incorporated in the owner's bid and offer. For instance, the owner bid only one of their units for pumping for five hours starting at $t=0$ and ending at $t=4$ indicated as the ``Pump Bid Window'' in Fig. \ref{fig:two_units_case_study},  such that if all their bids are cleared the reservoir would be charged to full at $t=4$.

The bid prices for pump loads submitted by the pumped-storage unit owners represent their willingness to purchase and consume the electricity in the pumping modes. The bid prices are calculated based on the owner's forecast of the system LMP and this forecast is unlikely to be accurate. In the example of current model shown in Fig. \ref{fig:two_units_case_study}, the bid prices for pump mode lead the system to clear the PSHUs to pump and fill the reservoir in the first five hours while losing the opportunity to pump in hour $5$ and hour $6$ with low generation costs and LMP. This myopic situation would impair the profits of the pumped-storage hydro unit owners and, in the meantime, the solution of the UC and ED problem deviates from a maximum social welfare. The myopic situation result from owner's bid and offer and the negative impacts on the system production cost in the current model will be further discussed in the MISO case study in section \ref{sec:MISOcase} with realistic scenarios.

\subsubsection{Proposed PSHU Model}
\label{cha:PSHU-proposed} In the proposed model, PSHUs do not submit offer/bid nor determine the scheduling window. Instead, they are fully optimized in the day-ahead market. State of charge of the reservoir is managed at the system level. In this study, the minimum and maximum energy capacity of the reservoir along with the state of charge of the reservoir at the beginning and the end time interval are listed in Table \ref{table:psh_reservoir}. The state of charge at the end of the day is required to return to the same level as the start, such that energy used for generation from the pumped-storage hydro units has to be stored earlier or recharged later within the studied time range. Noticed that the SOC limits listed in Table \ref{table:psh_reservoir} are explicitly modeled in the proposed model while those limits are implicitly incorporated in the owner's bid and offer in the current model as described in the previous section \ref{cha:PSHU-current-practice}. 

\begin{table}[!h]
	\caption{Reservoir}
	\label{table:psh_reservoir}
	\centering
	\begin{tabular}{c c c c c}
		\hline
		\multirow{2}{*}{}  & $\underline{E_r}$ & $\overline{E_r}$& $E_{r,1}$ & $E_{r,T+1}$  \\
		& MWh & MWh & MWh & MWh \\
		\hline
		Reservoir & 1000 & 3500 & 2600 & 2600\\
		\hline
	\end{tabular}
\end{table}

Thereafter, the bid price for pump loads is removed from the system objective. In addition, assume the generation costs for a PSHU is zero and ignore operation and maintenance costs and capital costs, the objective only contains the generation costs of the rest of the generators in the system besides the PSHUs as shown in \eqref{eq:objective_nobid}.


\begin{equation} \tag{\ref{eq:objective}'}
\label{eq:objective_nobid}
\underset{q,u,v,e}{\min} \sum_{g \in \cal{G}} \sum_{t \in \cal{T}}C(q_{gt}) 
\end{equation}

The PSHU solutions from the proposed model are indicated as red bars in Fig. \ref{fig:two_units_case_study}. Compare to the results from the current model, it can be observed that the same total amount of pumping energy is dispatched by the proposed model in the ``Dispatched Pump Window'' but the timing is shifted to when the system net-load is the lowest and accordingly with the lowest LMP. This improvement is beneficial in two perspectives. First, from the system operator point of view, overall system generation costs are reduced. The results in Fig. \ref{fig:two_units_case_study} demonstrate the load shifting effects of pump storage hydro units have been improved in the proposed model. PSHUs contribute more to reduce the system generation costs and to maximize the social welfare. Second, for PSHU owners, they are charged less for their pumping load with the proposed model. Given the same generation income in this case, the PSHU owner's profits are improved. A detailed cost benefit analysis of the proposed model in a realistic MISO system is provided in Section \ref{sec:MISOcase}.

\subsection{MISO Case Study}
\label{sec:MISOcase}

In this study we use a MISO case that includes 1,085 generators. Reserve requirements and transmission security constraints are included for all studies. Constraints on individual generators such as, minimum up/down time, maximum start up time, ramp constraints are included for all units including PSHUs with proposed model. We perform all tests on a $2.2$-GHz quad-core Intel Xeon CPU E5-2699 with $32$ GB Ram; all optimization problems are solved with Gurobi 8.0.

The proposed generic model in \eqref{eq:objective}-\eqref{eq:soct} has been implemented for a pumped storage hydro plant with multiple pumping and generating units in the MISO prototype day ahead security constrained unit commitment tool \cite{High2016pan}. The capacities of the unit for both pumping and generating are in GWs. An empirically hard to solve $36$-hour load scenario from MISO historical data library is selected for the computational study. There are three peaks in the scenario. The highest and lowest hourly demand are $91.6$ GW and $70$ GW respectively.


\subsubsection{Computational Study}
Table \ref{table:computational} shows the computational results of different models:

\begin{itemize}
\item Hippo: A High-Performance Power-Grid Optimization (Hippo) tool \cite{High2016pan}. MISO current pumped storage hydro model introduced in Section \ref{cha:PSHU-current-practice} is applied; 
\item Hippo + PSH: MISO current model is replaced by proposed pumped storage hydro model in Hippo;   
\item Hippo + CC: Hippo with configuration based combined cycle model \cite{dai2018configuration}.
\end{itemize}

The test case and load scenario with Hippo has been benchmarked with MISO production day-ahead market engine. Therefore Hippo is used as a benchmark for this study. As introduced in Section \ref{cha:PSHU-current-practice}, the PSHU currently offer costs and bid prices for their generation and pump mode respectively. State of charge of their reservoirs are not enforced by the system operator, instead, maximum daily generation is applied. Also, The pump/generate window for the unit are fixed. In Hippo + PSH, the particular PSHU is represented by the proposed model. In addition, a component-configuration based combined cycle model \cite{chen2017mip} that had been tested with Hippo is used for a comparison.  

\begin{table}[!h]
	\caption{Computational Results}
	\label{table:computational}
	\centering
	\begin{tabular}{c c c c c}
		\hline
		Model  & Hippo & Hippo + PSHU & Hippo + CC \\
		\hline
		Mip gap at 1200s & 0.11 \% & 0.16 \% & 0.28 \% \\
		\hline
	\end{tabular}
\end{table}

Table \ref{table:computational} shows the computational performance of all three MIPs at the cutoff time. The day ahead market cutoff solving time in MISO is $1200$ seconds. Solutions with MIP gap limit lower than $1\%$ will be accepted according to the MISO operating guide. Since the MIP gap of three models are below $1\%$, the results are considered acceptable in MISO operation. Because of the limited number of PSHUs and the proposed tight formulation, there is only a moderate increase in the computational difficulty compared to the benchmark Hippo model. 


\subsubsection{Benefit Analysis} In this section, the proposed model is benchmarked with the current model in examples based on real data in an actual day in MISO system. 

To make a fair comparison between the proposed model and the current model, given a start state of a reservoir and round-trip efficiency of the PSHU, the realized state of the reservoir at the last hour of the day from the results of the current model is applied to the proposed model. That is, the total energy charged to or discharged from the reservoir in each day in the simulation are the same for both models. To lay out a more realistic benefit analysis, the minimum SOC and maximum SOC of a reservoir in the proposed model is calculated from the results of current model. To be precise, the minimum SOC is calculated as the start state of the reservoir minus the effect of the generation cleared by the current model considering the efficiency. Similarly, the maximum SOC is the start state of the reservoir plus pumping cleared by the current model considering the efficiency. The PSHU parameters in the proposed model are summarized below.

\begin{itemize}
\item The SOC of a reservoir at the beginning of the day hour $0$ is given.
\item The SOC of a reservoir at the end of the day hour $24$ is calculated from the results of the current model and is fixed in the proposed model.
\item The SOC min and max are calculated from the results of the current model and are fixed in the proposed model.
\item All the other unit parameters (such as min up/ down time, ramp rate etc.) in the proposed model are copied from the production offer in the current model.
\end{itemize}

\begin{figure}
\centering
\includegraphics[scale=0.38]{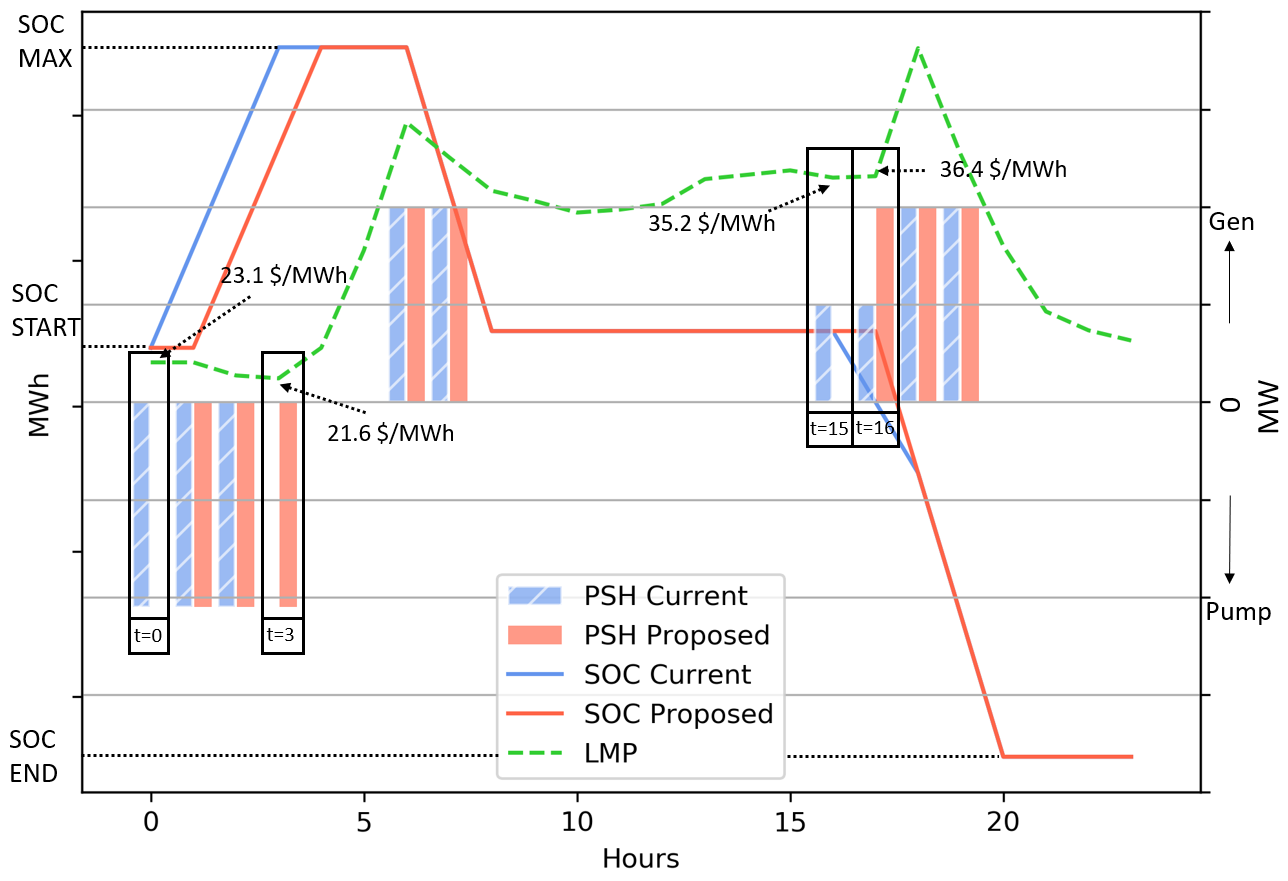}
\caption{MISO Case Study Results: PSHU $1$.}
\label{fig:MISO_Case_Study0}
\end{figure}

As shown in Fig. \ref{fig:MISO_Case_Study0}, the dispatch results of a PSHU in MW (aligned with the axis on the right side of the figure) from proposed model (solid red bars) are compared with current model (shaded blue bars). Notice that the positive side of the axis indicates generation and the negative side indicates pumping. The blue and red solid lines indicate the SOC of the current model and proposed model respectively with axis on the left using unit MWh. The dashed green line gives the trend of the Locational Marginal Price (LMP) at the PSHU connection node in the proposed model. 

Due to the tight SOC limits at PSHU $1$, the dispatch results of the unit in current model and proposed model are very close as shown in Fig. \ref{fig:MISO_Case_Study0}. In fact, the total amount of pumping and generation are the same in both models. However, the pump load and generation are allocated at slightly different periods in the day. The differences are highlighted in the black boxes with time indexes in Fig. \ref{fig:MISO_Case_Study0}. Comparing the proposed model to the current model, it can be observed that the same amount of pump load is shifted from $t=0$ to $t=3$ when LMP is lower. On the generation side, it can also be observed that some generation is shifted from $t=15$ to $t=16$ when LMP is higher. Although the differences between the current model and the proposed model are small for PSHU $1$, this indeed demonstrate the centralized optimal effects of the proposed model against the current model in a very precise and concrete way. As the results show, the proposed model would find the optimal solution that allocates the pump load to the intervals when the system is least stressed as indicated by the lower LMPs and allocates the generation to the intervals when the system is most stressed as indicated by the higher LMPs.

\begin{figure}
\centering
\includegraphics[scale=0.38]{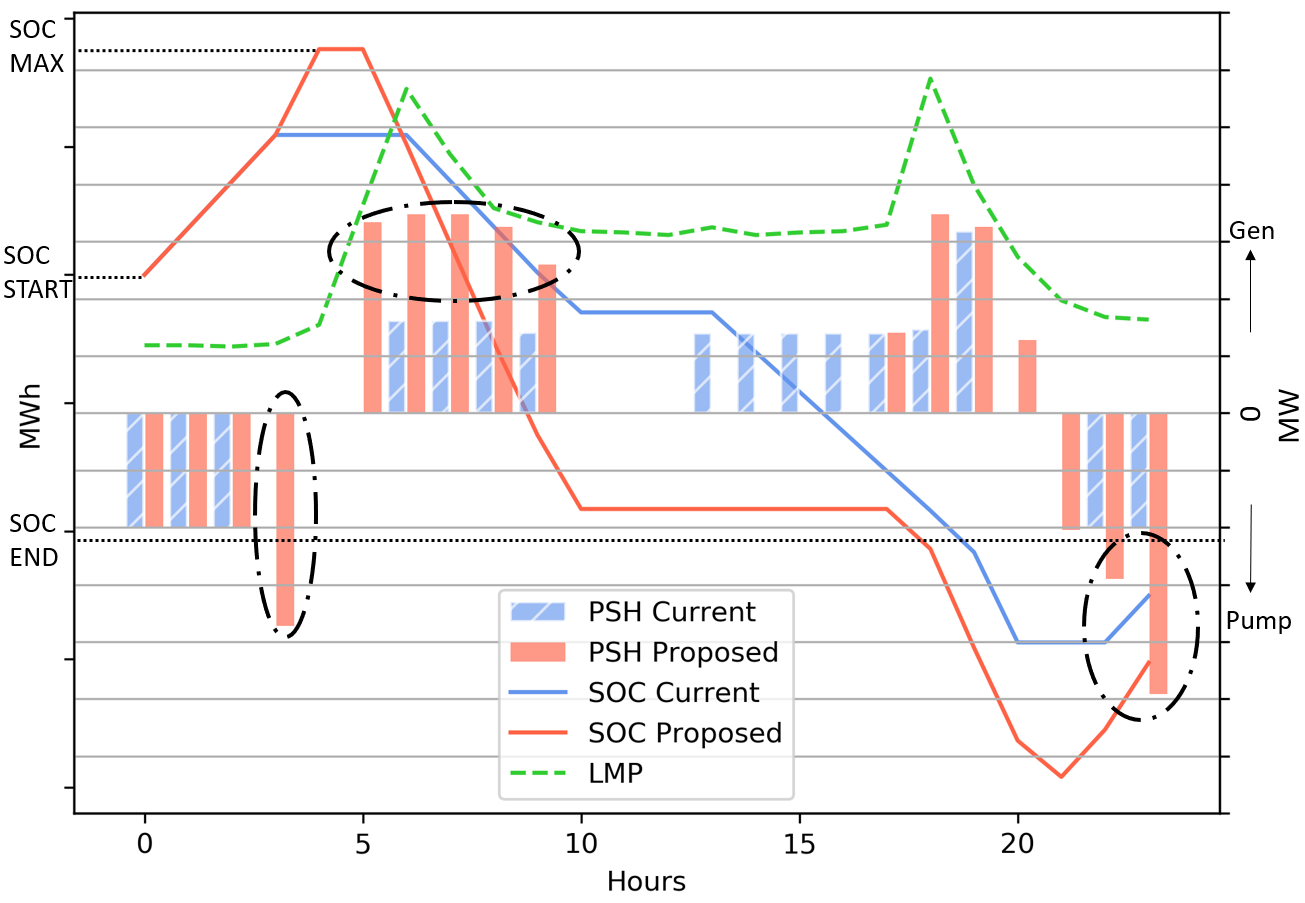}
\caption{MISO Case Study Results: PSHU $2$.}
\label{fig:MISO_Case_Study1}
\end{figure}

The same study has been applied to a different PSHU $2$, as the results are shown in Fig. \ref{fig:MISO_Case_Study1}. The PSHU parameters of the unit in the proposed model are estimated in the same way from current model. The notations in Fig. \ref{fig:MISO_Case_Study1} are exactly the same with Fig. \ref{fig:MISO_Case_Study0}. The key observation from Fig. \ref{fig:MISO_Case_Study1} is, due to the fact that PSHU $2$ has a larger reservoir and larger generate/pump capacities than PSHU $1$, the proposed model would shift more pump loads and generations to optimal positions highlighted in the black dashed ellipses. Compared to results in Fig. \ref{fig:MISO_Case_Study0}, the benefits of the proposed model are more significant for the PSHU with a larger reservoir and larger generation and pump capacities.

To briefly sum the qualitative benefits, especially shown in Fig. \ref{fig:MISO_Case_Study1}, the PSHU becomes more active and flexible with the proposed model compared to the current model. From the results of both units in Fig. \ref{fig:MISO_Case_Study0} and Fig. \ref{fig:MISO_Case_Study1}, the pumping and generation from proposed model follows the system conditions closer than the current model, that is, with the proposed model, the PSHU pumps more when the LMP is low and generate more when the LMP is high. In the current model, a PSHU is dispatched based on the bid and offer from the unit owner. In particular, pumping is restricted to specific hours based on the owner forecast of minimum prices, but the minimum prices often occur at different times to that forecast. Inherently because a unit owner has limited system information, the PSHU owner can not perform better than the system operator on forecasting the system conditions in the Day-ahead Market. This gap between the current model and the proposed model would be enlarged if the system become more dynamic and even harder for the unit owner to predict. For example, the flexibility of the unit and the system optimality from proposed model will be more significant when more renewable especially solar generations are included in the system.

\begin{table}[!h]
	\caption{Benefit Analysis}
	\label{table:benefit analysis}
	\centering
	\begin{tabular}{c c c c c c}
		\hline
		\multirow{2}{*}{}  & System & PSHU 1 & PSHU 2 & PSHU 3\\
		& Objective [\$] & Profit [\$] & Profit [\$] & Profit [\$]\\
		\hline
		Improvement \% & $0.4\%$ & $1\%$ & $10.8\%$ & $6\%$\\
		\hline
	\end{tabular}
\end{table}

The benefits of the proposed model demonstrated in Fig. \ref{fig:MISO_Case_Study0} and Fig. \ref{fig:MISO_Case_Study1} are quantified and summarized in Table \ref{table:benefit analysis}. The unit commitment solutions of the rest generation units in the system other than the three PSHU from the current model are fixed to the proposed model, such that the results shown in Table \ref{table:benefit analysis} mainly reflects the impacts of the proposed model on the PSHUs. The reduction in system objective from the proposed model is shown as the percentage of the system objective of the current model. 
At the same time, the profit increment for the PSHU owners from the proposed model are shown as percentages of their profits result from current model. PSHU $2$ and $3$ have larger reservoirs and larger generate/pump capacities, therefore the proposed model gained more significant improvement compare to PSHU $1$. A more realistic calibration of benefits can be done by considering both day ahead and real time market. However, the results in this study demonstrate that the proposed model would improve both social welfare and PSHU owner's profits assuming perfect forecast in day ahead market. We will leave the multi-stage market study to the future work. 

In addition, the scenario used in this study is from historical data library and has less renewable interconnections. According to current MISO generation interconnection queue \cite{GIO2019miso}, significant amount of renewable units are likely to be interconnected in the near future. In a system with more variations and intermittencies, the value of the flexibility from a PSHU is expected to be further escalated with the proposed model.

\section{Conclusion and Future Work}
\label{sec:conclusion}
The current model for PSHU in MISO day-ahead market is rigid and leads to less economic dispatch solutions. In this study, we propose a new formulation of PSHU that enhances the flexible dispatch of the unit in day-ahead market. A MISO case study shows the proposed model improves system welfare and increases PSHU owners' profits.

Although this paper focuses on the modeling of a PSHU in a day ahead market, the proposed configuration based model can be applied to other storage resources with multiple operating modes and states of charge. Also, by specifically modeling the transitions, the proposed model can be easily extended to include detailed operational features during transitions in a PSHU in a market with smaller time intervals. The numerical results show the applicability and scalability are promising for large systems.


%


\section*{Acknowledgment}
This material is based upon work supported, in part, by the U.S. Department of Energy’s Office of Energy Efficiency and Renewable Energy (EERE) under the Water Power Technologies Office Award Number DE-EE0008781.



\ifCLASSOPTIONcaptionsoff
  \newpage
\fi



%

\bibliographystyle{IEEEtran}
\bibliography{Pumped_storage.bib}

%








\end{document}